\documentclass{aa}
\usepackage{graphicx}
\usepackage{natbib}
\begin{document}
\title{The photospheric environment of a solar pore with light bridge}
\author{S. Giordano, F. Berrilli, D. Del Moro \and V. Penza}
\offprints{francesco.berrilli@roma2.infn.it}
\institute{Department of Physics, University of Rome ``Tor Vergata'', I-00133, Rome, Italy}
\date{Received , 2007; accepted XX, YY}
\abstract
{
Pores are one of the various features forming in the photosphere by the emergence of magnetic field onto the solar surface.
They lie at the border between tiny magnetic 
elements and larger sunspots.
Light bridges, in such structures, are bright features separating umbral areas in two or more irregular regions.
Commonly, light bridges indicate that a the merging of magnetic regions or, conversely, the breakup of the area is underway: 
}
{
We investigate the velocity structure of a solar pore (AR10812) with light bridge, and of the quiet solar photosphere nearby, analyzing high spatial and spectral resolution images.
}
{
The pore area has been observed with the Interferometric BI-dimensional Spectrometer (IBIS) at the Dunn Solar 
Telescope, acquiring monochromatic images in the Ca~II 854.2 nm line and in the Fe~I 709.0 nm line 
as well as G-band and broad-band images.
We also computed the Line of Sight (LoS) velocity field associated to the Fe~I and Fe~II photospheric lines.
}
{
The amplitude of the LoS velocity fluctuations, inside the pore, is smaller than that observed in the quiet granulation near the active region.
We computed the azimuthal average LoS velocity and derived its radial profile.
The whole pore is characterized by a downward velocity $\simeq-200~m\cdot s^{-1}$ and by an annular downflow 
structure with an average velocity of $\simeq-350~m\cdot s^{-1}$ with respect to the nearby quiet sun.\\
The light bridge inside the pore, when observed in the broad-band channel of IBIS and in the red wing of Ca~II 854.2 nm
 line, shows an elongated dark structure running along its axis, that we explain with a semi-analytical model.
In the highest resolution LoS velocity images the light bridge shows a profile consistent with a convective roll: a 
weak upflow, $\sim 50\div 100~m\cdot s^{-1}$, in correspondence of the dark lane, flanked by a downflow, 
$\sim -(200\div 300)~m\cdot s^{-1}$.
}
{}
\keywords{Physical processes: convection - Sun: photosphere - Sun: sunspots}
\titlerunning{The photospheric environment of a pore}
\authorrunning{Giordano et al.}
\maketitle
\section{Introduction}
The inhomogeneous and structured aspect of the solar photosphere essentially originates from convective flows, 
carrying energy from the deeper layers of the star, and from the magnetic field, emerging at the surface. 
The photosphere shows a wide variety of magnetic features, ranging from the largest sunspots, with typical field 
strengths of $\simeq 3000$~G, down to the 0.1~Mm scale magnetic elements, with typical field strengths of $\simeq 1500$~G. 
In this family of solar magnetic structures, pores represent the link between tiny flux tubes and complex and large sunspots.
The pores do not present a penumbra and are small (1-6~Mm) and intense concentration of magnetic field, typically 
$\simeq 1700$~G.
For a review about pores and sunspots see \citet{sobotka03} and \citet{TW}.\\
Across a pore, the magnetic field strength exhibits a variation from 600~G to 1700~G, as reported by \citet{sutterlin}.
This behavior was confirmed by \citet{keppens} who found a decrease of the vertical magnetic field component from 
1700~G, in the pore center, to 900~G at its magnetic edge.
More in detail, magnetic field lines are found to be roughly vertical in the center of pores, while they are inclined 
by about $40^\circ -80^\circ $ at their boundaries.\\
Observations show that pore and sunspot umbrae exhibit an inner structure.
The umbra contains a large variety of fine bright features, like umbral dots or light bridges, a hint of a convective 
energy transfer.
The occurrence of a structured umbra is theoretically accounted by two categories of models: a monolithic and 
inhomogeneous flux tube (e.g. \citet{choudhuri}) or a cluster of single flux tubes (e.g. \citet{parker}).
Both models, although starting from different assumptions, predict the presence of fine and bright features 
embedded in the dark umbra.
For the monolithic model, observed fine features are related to convective motions not completely inhibited in 
the sub-photospheric layers.
For the cluster model, bright structures are explained as signatures of field-free gas plumes penetrating from 
below into the photosphere.\\
Light bridges (hereafter LB) are bright irregular elongated features crossing the umbra of sunspots and pores \citep{sobotka03,TW}, manifesting a great range of variability in their morphology and physical properties.
A first interpretation of photospheric LB came from \citet{vasquez73}, who accounted for them as the result of sunspot decay preceding the restoration of the granular surface.
Instead, \citet{rimmele} measured a positive correlation between their brightness and upflow velocities, which was explained as an evidence of a magneto-convective origin.
These findings supported what was already suggested by \citet{hirzberger} in their study of the evolution of the small bright grains forming a LB.
\\
Concerning the formation of pores, observations (e.g. \cite{wang}, \cite{keil})  reveal that pores result from the merging
of small magnetic elements, driven by supergranular and subsurface flows. After the pore has formed, it can evolve into a sunspot if 
the magnetic flux increases and the magnetic field becomes more inclined at the edge of the pore.
The possible evolution of pores into sunspots depends on a dynamic stability criterion \citep{bray,wang,rucklidge}.
According to the model of \citet{rucklidge}, there exists a critical magnetic flux, below which the pore size can grow without becoming a sunspot.
This model also accounts for the overlap in sizes between small sunspots and larger pores.\\
Several observational studies reported the presence of an annular structure of strong downflows around the pore (e.g.\citet{keil,san}) and the possible presence of supersonic 
speeds near the edge \citep{uit}. Such downflows around magnetic elements and pores are predicted also by two-dimensional MHD simulations and 
magnetoconvection (e.g. \citet{steiner, hulburt}).\\ 
Several analysis of photometric data (e.g. \citet{rbs}) showed that the horizontal flows around a pore are moving towards and across its umbral boundaries.\\
The aim of this paper is to investigate the photospheric environment of a roundish pore with a light bridge supposedly formed by the quasi-merging of two separate dark structures.
More in detail, we concentrate our study on the evolution of its LoS velocities, mean radial structure and dynamics.
We also investigated the dynamics and the brightness profile of the light bridge.
The previous history of the pore, classified as AR10812, is derived from MDI/SOHO magnetograms and continuum images.\\
The paper is structured as follows: in \S~2 we give a brief account of the observations and the calibration 
procedure; in \S~3 we discuss the synthesis and Velocity Response Function of IBIS lines and the procedure to compute 
the LoS velocities; in \S~4 we describe the physical properties of the pore and of the light bridge, whose intensity 
behavior we explain through a simple semi-analytical model. In \S~5 we summarize our findings and propose a sketch of the pore configuration.
   
\section{Observations and data processing}
\subsection{Observations}

\begin{figure}
   \centering\includegraphics[width=9cm]{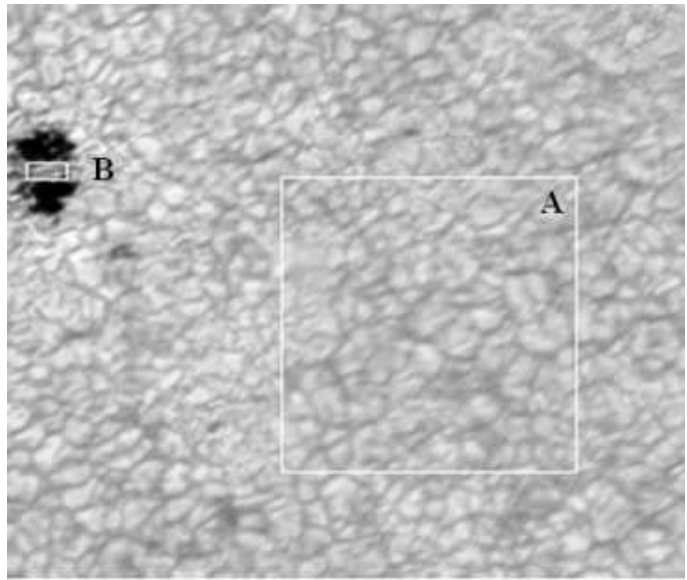}
      \caption{Broad-band image of the FoV, including the pore, after applying  a 
	  speckle restoring program. We outline the analyzed regions: the quiet granular field ($A$) and the light bridge ($B$).}
         \label{figure1}
   \end{figure}
The observations were performed on September, 28th 2005 at the 0.76 m DST in Sunspot, New Mexico. We observed a central region 
of the solar disk including a pore with light bridge (AR10812).
The interferometer used for these observations was the Interferometric BI-dimensional Spectrometer (IBIS) 
\citep{cavallini}, fed by the High Order Adaptive Optics (HOAO) system \citep{rimmele} tracking the pore.\\
The IBIS instrument allows us to obtain solar monochromatic images with high spatial ($\simeq$ 0.2 arcsec), 
spectral ($\lambda/\Delta\lambda=2\cdot 10^5$) and temporal resolution (exposure time $\simeq 10~ms$; acquisition rate 
$\simeq 5$~frames$\cdot$ $s^{-1}$). 
IBIS essentially is formed by two air-spaced Fabry Perot Interferometers (FPIs), 50 mm in diameter, used in 
classical mount and in axial mode, in series with one of five prefilters with FWHM of 0.3 nm to 0.5 nm, 
depending on wavelength, mounted on a filterwheel.
Monochromatic images, consisting of 80" diameter circular section, are recorded by a Princeton CCD camera. 
The detector is a Kodak KAF-1400 with 1317 $\times$ 1035 pixels, a dynamic range of 12 bits and an acquisition rate 
of 5~Mpixels$\cdot s^{-1}$.
IBIS is equipped with a white light channel, that provides broad-band images with a 5.0 nm passband around 700.0 nm, 
strictly simultaneous to the narrow-band ones.\\
The dataset used in this paper consists of 200 sequences, containing a 16 image scan of the Fe~I 709.0 nm line, 
a 14 image scan of the Fe~II 722.4 nm line and 6 spectral images of the Ca~II 854.2 nm line (one line core image 
and 5 line wing images).
The exposure time for each monochromatic image was 25 ms.
The CCD camera was rebinned to 512 $\times 512$ pixels, so that the final pixel scale for the images
was $0.17"\cdot$pixel$^{-1}$.
The time interval between two successive images and two successive spectral series was 0.3~s and 14~s, respectively.\\
In addition to the narrow-band images, G-band images were simultaneously recorded.
\subsection{Standard reduction of IBIS data}
The first step in the data reduction is to correct both the data and the flatfield images for dark current and CCD non-linearity effects.
Regarding the flatfield correction, we have to consider the blueshift effect due to the classical mounting of the two FPIs.
In this configuration every image point corresponds to rays with a specific angle with respect to the optical axis propagating through the FPIs.
This causes a systematic blueshift of the instrumental profile when moving from the optical axis towards the edge of the FoV, therefore reaching its maximum in the outermost pixels.
This maximum is about 0.6 nm at 600.0 nm wavelength and is about 1.0 nm at 850.0 nm wavelength.\\
To produce the gain table, we compute a ``flatfield scan'' averaging the 100 flatfield sequences for each spectral point.
We then determine the instrumental blueshift map by calculating the line core shifts of all the pixels in the FoV with respect to the reference profile and fitting this resulting map with a parabolic surface.

The reference profile is obtained by averaging the line profiles of the pixels in a central region (100 $\times$ 100 pixels) of the FoV.
In order to obtain an average spectral profile, all pixel profiles are remapped to a common wavelength scale by applying the blueshift map and then averaged.
The remapping process is performed by linearly interpolating the measured spectral profiles.
The average spectral profile is then applied on each pixel and shifted accordingly to the blueshift map in order to construct the ideal blueshifted flat field scan that is expected 
for a ``perfect'' system (i.e. with only the blueshift contribute and no gain variation).
Any differences between the flat field scan and the ideal blueshifted flat field scan are due to spatial inhomogeneities in the system response.
The final gain table scan is constructed by dividing the flat field scan by the ideal blueshifted flat field scan, so that it does not contain unwanted spectral information.
The gain table is then applied to the raw data in order to correct each pixel for the flat field response.
Finally, the blueshift correction for the flat fielded data is computed and applied with the same process used for the flat fields.\\
\subsection{Key properties of IBIS photospheric and chromospheric lines}
\begin{figure}
   \centering\includegraphics[width=9cm]{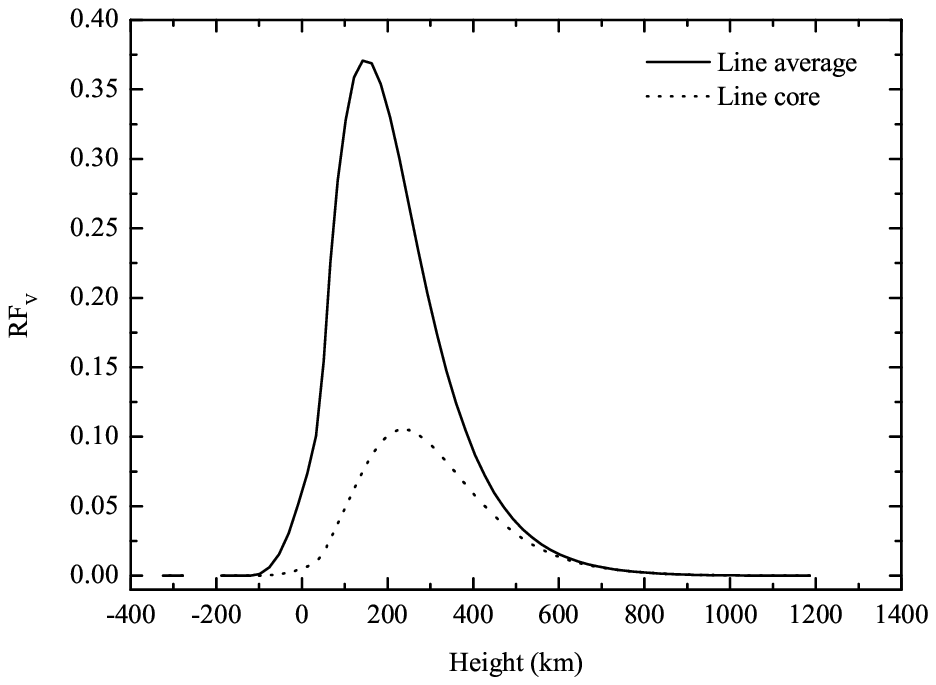}
      \caption{Velocity response function for the Fe~I 709.0 nm line. $Solid$ $line$: average $RF_{V}$; $Dotted$ $line$:
	  center line position $\lambda _{0}$.}
         \label{figure2}
   \end{figure}

\begin{figure}
   \centering\includegraphics[width=9cm]{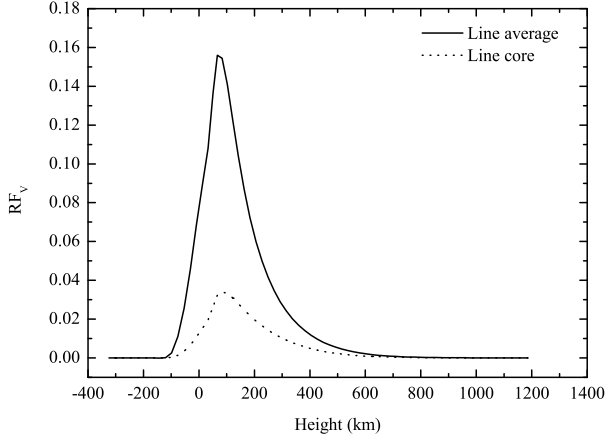}
      \caption{Velocity response function for the Fe~II 722.4 nm line. $Solid$ $line$: average $RF_{V}$; $Dotted$ $line$:
	  center line position $\lambda _{0}$.}
         \label{figure3}
   \end{figure}

To associate to observed photospheric lines a suitable ``formation zone'', we study their sensitivity, as a function of
 the depth, to the perturbations of velocity.
In detail, we treat the effects of linear dynamic perturbations on the line profiles and study the velocity response 
functions $RF_{\,V}$ of the emergent intensity at the observed wavelengths within the lines \citep{caccin77,berrilli02},
 that provide the corresponding intensity perturbation as:
\begin{equation}
 \label{deltaI}
\delta I(\lambda) = \int_{-\infty}^{+\infty} v(z) RF_V(z,\lambda) dz.
\end{equation}
The velocity response function is computed by using the usual formula:
\begin{equation}
 \label{RF}
RF_V(z,\lambda) = \frac{\partial \chi(z,\lambda,v)}{\partial v} (S(z,\lambda) - I(z,\lambda)) e^{-\tau(z,\lambda)}.
\end{equation}
where $\chi(z,\lambda,v)$ is the total opacity for volume unit, $S(z,\lambda)$ the source function and $\tau(z,\lambda)$ 
the optical depth.
\begin{table}   
$$
\begin{array}{l l c c r c}
\hline
\noalign{\smallskip}
$$ Line & \chi_{ion}\,(eV)& \chi_{ex}\,(eV)& \lg(gf)& \zeta\,\,\, &\xi\,(km/s)$$  \\
\noalign{\smallskip}
\hline
\noalign{\smallskip}
$$ FeI\phantom{1} \, 709.0 \, \phantom{1}nm & \phantom{1}7.87 & 4.231 & -1.3 & 17 & 1.00 $$ \\            
$$ FeII \, 722.4 \, \phantom{1}nm & 16.18 & 3.889 & -3.3 & 50 & 1.75 $$ \\
\noalign{\smallskip}
\hline
\end{array}
$$
\caption[]{Line parameters used for the calculation of line profiles. $\zeta$ is a fudge factor multiplying the Uns\"{o}ld value of $\gamma$ in the Lorentz part of the absorption profile, while $\xi$ is the usual microturbulence term.
The values of $\zeta$ and $\xi$ are chosen to optimize the comparison with atlas data.}
\label{linepar}
\end{table}
For the calculation we adopted LTE approximation, so that the $S(z,\lambda)$ is velocity independent, 
and we use Kurucz's solar atmospheric model \citep{kur94}.
The validity of this model and of the line parameters we used (Table \ref{linepar}) is given by the comparison 
between the theoretical synthesis of the lines and the atlas data \citep{kurATLAS}.\\
The formation depth, i.e. the location of the maximum of the $RF_V$, depends on the wavelength, hence on the given  
points of the line profile.
This dependence is more evident for the Fe~I 709.0 nm line (Fig. \ref{figure2}),  while for Fe~II 722.4 nm the 
$RF_V$ maximum is almost the same for each wavelength (Fig. \ref{figure3}).
Since we derive the velocity shift by a fitting procedure that uses all the sperimental spectral points, 
we consider the mean of the corresponding $RF_V$.
The formation heights for the average lines are reported in Table \ref{Hformation}.\\
The Ca~II 854.2 nm chromospheric line forms at $\simeq 800$ km above the photosphere. 
Our observations, on the red wing of this line ($\lambda_0$ + 12 nm), may be associated to a photospheric  height, as indicated
by the Contribution Functions computed for a quiet Sun model \citep{QuXu02}. 
This estimation is an approximation as the region we observe is deeply immersed in a non-quiet and non-homogeneous atmosphere.
 
\begin{table}   
$$
\begin{array}{l l l l}
\hline
\noalign{\smallskip}
$$ Line    & h_{core}\,\, (km) & h_{mean}\,\, (km)  & FWHM $$  \\
\noalign{\smallskip}
\hline
\noalign{\smallskip}
$$ FeI\phantom{1}\, 709.0 \phantom{1}nm   & 150  & 100 & 200  $$ \\
$$ FeII\,722.4 \phantom{1}nm   &  \phantom{1}40  & \phantom{1}40  & 140 $$ \\
\hline
\end{array}
$$
\caption[]{Line formation depth for the line center and the whole line. The latter is obtained by averaging the different $RF_V$ computed in the sperimental wavelenghts and its FWHM is also given.}
\label{Hformation}
        \end{table}

\subsection{LOS Velocity maps}
Vertical velocity maps were computed for the Fe~I 709.0 nm and Fe~II 722.4 nm lines by applying a line-profile 
Gaussian fit to the monochromatic cube of images and then transforming Doppler shifts in LoS velocities.
The same procedure has been used to compute spectral line core intensity and FWHM maps.\\
In order to study the convection dynamics, before analyzing velocity and intensity fields,
we have to take into account the 5-min acoustic oscillations.
These are removed by applying a $k_{h}-\omega$ subsonic filter: we cut a cone out of the $k_{h}-\omega$ space 
whose outer borders correspond to $7~km~\cdot s^{-1}$.\\
After the application of the subsonic $k_{h}-\omega$ filter on the time series of continuum, velocity and center-line 
intensity maps, we consider a region of interest (Fig. \ref{figure1}) centred on the pore of about 27''$\times$27''.\\
This is the region tracked by the adaptive optic system, so we are confident that it is characterized by a high spatial 
resolution.\\
In our analysis, we neglect the velocity offset due to the convective blueshift \citep{keil} and define absolute 
values by setting to zero the mean velocity of the whole observed field. 

\section{Results and discussion}
In order to study the evolution of AR10812, we examined high resolution MDI continuum images, corresponding to three 
days before our observation run, and established that it was made up of two different structures, showing the same polarity.
If we look at AR10812 three days after our observation date, we see that the photospheric and magnetic signatures have disappeared.

\subsection{Radial structure of the pore}
\begin{figure}

   \centering\includegraphics[width=9.5cm]{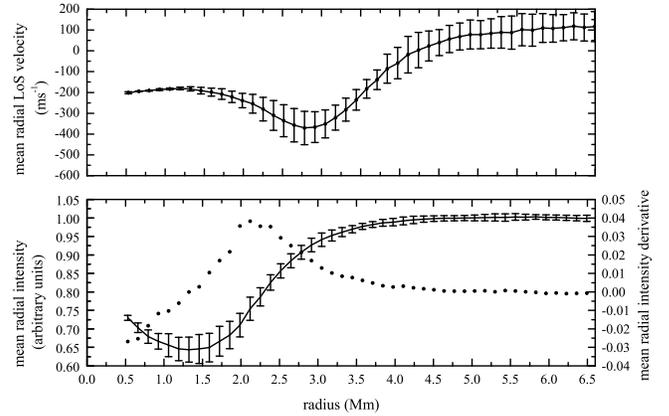}
      \caption{{\it Upper panel}: plot of the mean radial LOS velocity inside the pore; {\it lower panel}: plot of the mean 
	  radial intensity ({\it solid}) and of its derivative ({\it dotted}).} 
         \label{figure4}
   \end{figure}
\begin{figure}
\centering\includegraphics[width=8.5cm]{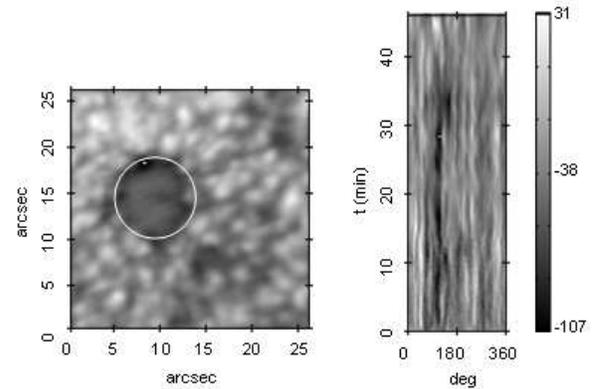}
\caption{$Left$ $panel$: time averaged Doppler map of the analyzed pore region; $right$ $panel$: time-slice of the 
circular ring (white circle) surrounding the umbra. 
It is obtained by following the evolution of the structures covered by the circle surrounding the pore. 
The center of our polar coordinates system corresponds to the center of the pore.}
\label{figure5}
\end{figure}

We investigated the radial structure of the circular pore, by computing the azimuthal average LoS velocity and intensity excluding the 
light bridge contribution. 
We focus our analysis on the Fe~I 709.0 nm LoS velocity maps. By analyzing the plot shown in the upper panel of Fig. \ref{figure4} 
it is worth noting that an annular downflow lies just outside the pore border, as 
identified by the maximum in the derivative of the intensity.
From a qualitative point of view the LoS velocity shows three different behaviors: it results negative and 
quasi-constant inside the umbra; it reaches its largest negative values, $\sim-500 m\cdot s^{-1}$, just outside 
the umbra; beyond this downflow region the velocity increases toward typical granular values.\\
A more careful investigation of the annular downflow region points out an irregular form in space and an 
intermittent behavior in time. As a matter of fact, a time-slice representation (Fig. \ref{figure5}, $right$ $panel$) 
of this downflow region shows that recurrent strong downflows are present in the upper-left boundary of the pore, 
this resulting in persistent downflows in the whole period averaged LoS velocity map (Fig. \ref{figure5}, $left$ $panel$).

\begin{figure*}[t]
   \centering\includegraphics[width=16cm]{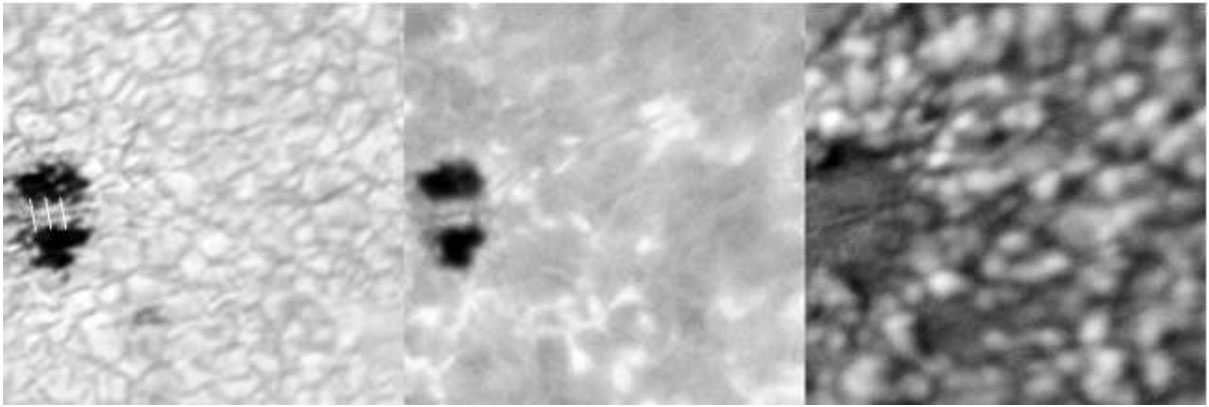}
      \caption{The (AR10812) pore region intensity, as observed with the broad-band channel of IBIS {(\it left panel)} 
	  and on the red wing of Ca~II~854.2~nm line (central panel). The corresponding LoS velocity pattern, computed 
	  from Doppler shifts of Fe~I~709.0~nm line, is shown in the right panel, a bright (upward) LoS velocity feature 
	  is clear visible along the light bridge axis.
 The white segments orthogonal to the light bridge, drawn in the
	  left panel figure, denote the cuts across which the intensity and velocity profiles plotted in Fig. \ref{figure7} have been computed.}
         \label{figure6}
   \end{figure*}
   
\subsection{Intensity and velocity structure inside the pore}
\begin{figure}[t]
\centering\includegraphics[width=9cm,height=9cm]{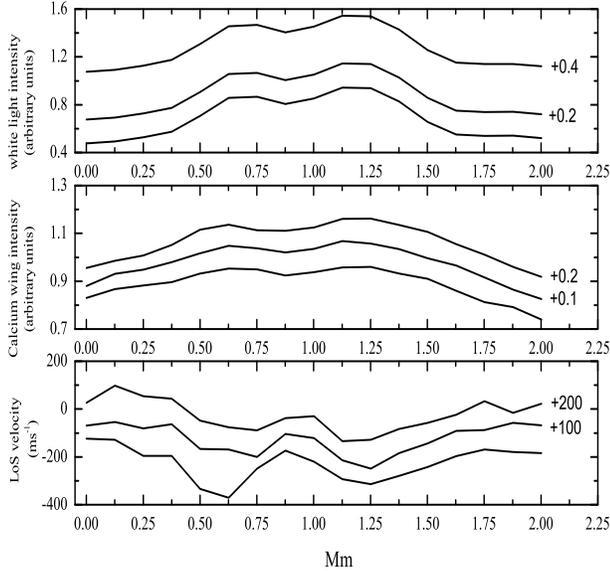}
\caption{Plots of the light bridge intensity, in the broad-band channel of IBIS ({\it upper panel}) and in the red wing of the Ca II
854.2 nm line ({\it central panel}) and of the corresponding Fe~I~709.0~nm line LoS  velocity ({\it lower panel}). All profiles have been computed along the 
segments shown in Fig. \ref{figure6} and have been shifted vertically by arbitrary amounts to provide separation.}
\label{figure7}
\end{figure}

The dynamics of the light bridge are investigated by means of intensity and LoS velocity maps. Fig. \ref{figure6} 
shows the pore region intensity, as observed with the broad-band channel of IBIS and on the red wing of 
CaII~854.2~nm line, and the corresponding LoS velocity pattern, computed from Doppler shifts of Fe~I~709.0~nm line. 
The intensity and velocity profiles, along slices which are orthogonal to the light bridge, are shown in Fig. \ref{figure7}.
\\
From our analysis we may distinguish two 
major outcomes: the presence of elongated features, both in intensity and in LoS velocity images, and the 
occurrence of a kind of reversing in intensity and velocity of small scale features, i.e. intensity maxima along the light bridge
match more intense downward velocities.\\
The observed reversing in the LB intensity and velocity frames, recalls the {\it inverse granulation} phenomenon, 
which consists of the inversion of temperature fluctuations, with respect to velocity field, in the upper photosphere. 
The occurrence of reversed granulation, around 120~km above quiet Sun photosphere, was observed at disk center in 
Fe~I~537.9~nm and Fe~I~557.6~nm photospheric lines by THEMIS in IPM imaging mode \citep{berrilli02}. Successively, 
\cite{Puschmann03} reported an inversion of temperature, at a height of $\simeq 140~km$, using one-dimensional 
slit-spectrograms taken in quiet sun. More recently, \cite{janssen06} confirmed that reversed granulation is also 
visible at $\simeq 200~km$ using FeI~709.0~nm line center images. However, the inverse granulation phenomenon involves 
the upper quiet photosphere, whereas dark intensity features in broad-band channel images refer to a zero altitude 
photosphere. Moreover, our observations show that an elongate structure along the axis of the light bridge exists 
also in velocity maps. More in detail, our highest resolution images show that a weak upflow, $\sim 50\div 100~m\cdot s^{-1}$,
is present along the light bridge axis, while a downflow, $\sim-(200\div 300)~m\cdot s^{-1}$, exists along the boundary. 
The topology of such a velocity structure resembles some roll features known from laboratory experiments on 
Rayleigh-B$\acute{e}$nard convection and may be a signature of modified photospheric convective flows confined by two 
magnetic walls. 

\subsection{A light bridge dark lane semi-analytical model}

\begin{figure}
\centering\includegraphics[width=9cm]{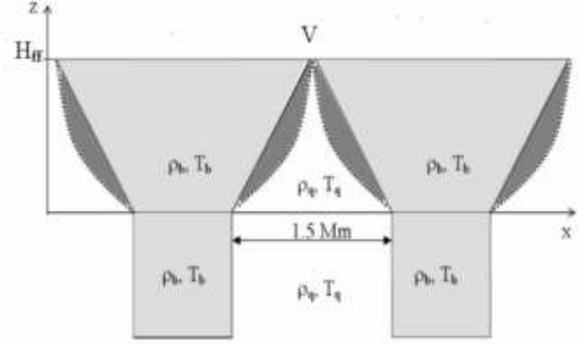}
\caption{Cross-section of the light bridge structure modeled as a field free plasma ({\it white region}) trapped by two magnetic flux tubes ({\it grey regions}). 
Two different geometries are considered for the surfaces separating the quiet and magnetic atmospheres: a straight line ({\it bold}) or a cubic function ({\it dotted}).} 
\label{figure8}
\end{figure}

\begin{figure}[h]
\centering\includegraphics[width=9cm,height=9cm]{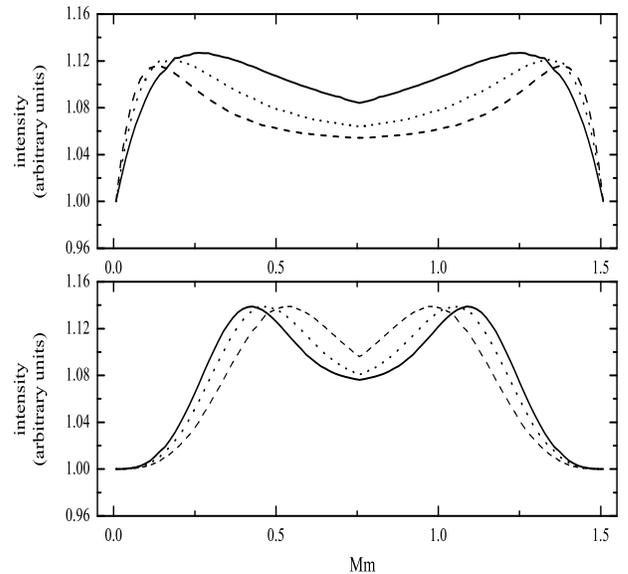}
\caption{Synthetic emergent intensities across the light bridge for the two different 
geometries of the separating surface: a straight line ({\it upper panel}) or a cubic function ({\it lower panel}). In both figures the 
three plots correspond to different $H_{ff}$ values (where $H_{ff}$ is the height of the V point shown in Fig \ref{figure8}): 200 km, 300 km and 400 km for solid, dotted and dashed lines, 
respectively.}
\label{figure9}
\end{figure}

The presence of a narrow central dark lane running along the axis of the light bridges is a common feature of LBs. 
Its existence is reported in \cite{sobotka94} and with more details in \citep{berber,litesetal}. Evidence of a 
magnetoconvective origin for a sunspot light bridge is reported by \cite{rimmele97}.
Recently, \cite{spruit} argued that a 3D radiative magnetohydrodynamic Nordlund \& Stein  simulation may explain the 
formation of a dark lane inside the light bridge as a consequence of the higher gas pressure in the field-free part of the 
photosphere, trapped between two magnetic fields.\\
To reproduce this configuration, we developed a simple thermal model 
of the light bridge, in which we consider a quiet, field free model below a magnetic zone partially emptied of plasma (see Fig. \ref{figure8}). \\
The thermal quantities, relative to the quiet sun (field-free), are given by the following analytic expressions, that 
well fit the solar Kurucz model \citep{kur94}:
\begin{eqnarray*}
\rho_q(\zeta) &=& \rho_q \frac{e^{z/H}}{e^{z/H}+1} = \rho_0 \frac{\zeta-1}{\zeta} \\
P_q(\zeta) &=& P_q ln(\zeta)	\\
T_q(\zeta) &=& T_q \frac{\zeta}{\zeta - 1}ln(\zeta) \\
\end{eqnarray*}
where $\zeta=e^{z/H}+1$ and z is the depth. This model satisfies the hydrostatic equation ($P_q=g \rho_q H$) 
and the perfect gas 
law ($T_q = \mu P_q/R \rho_q$). The two parameters H and $\rho_q$ are fixed so as to optimize the comparison with 
the Kurucz model ($\rho_q= 5.007\cdot 10^{-7}~gr\cdot cm^{-3}$ and $H=100~km$).  \\
For the magnetic atmosphere, we mimic the presence of the magnetic field 
through a model with an exponential density law:
\begin{eqnarray*}
 \label{b_model}
\rho_b(\zeta) &=& \rho_b (\zeta - 1)^\alpha \\
P_b(\zeta) &=& P_b (\zeta - 1)^\alpha \\ 
T_b(\zeta) &=& T_b\\ 
\end{eqnarray*}
where $\alpha$ is the ratio between the scale height of quiet and magnetised model ($\alpha=H/H_b$, $P_b=g \rho_b H_b$ 
and  $T_b = \mu/R g H_b$). In order to calculate the emergent intensity we need an expression for the opacity. We use the 
usual formula: $k(P,T)\propto P^A T^B$, where $A=0.5$ and $B=8$. These values are obtained through the best fit of the opacity 
tables computed by \cite {kur94}.
The $H_b$ parameter is a free parameter set to 120 km in order to match the observed contrast ($\simeq 1.5$) 
between umbra and quiet granular field. 
Eventually, the model (Fig. \ref {figure8})  uses two atmospheres (field-free and magnetic) to create a simplified geometry above the light bridge.
The geometry of the separating surface is described by two functions: a linear function ($z_t=a \cdot x$) and a cubic function ($z_t=b \cdot x^3$), where x is 
the horizontal coordinate. We adopt a quiet model for $z > z_t$ and a magnetic model for $z < z_t$.
Considering the observations, we fix the light bridge width to 1500 km. In Fig. \ref{figure9} we show the intensity profile 
inside the light bridge for the two adopted geometries. It's worth to note that both geometries are able to qualitatively
reproduce the presence of a dark lane across the light bridge, but with the cubic function we obtain a contrast profile, that better 
matches the observed one. 

\section{Conclusions}
\begin{figure}[t]
\centering\includegraphics[width=7.5cm]{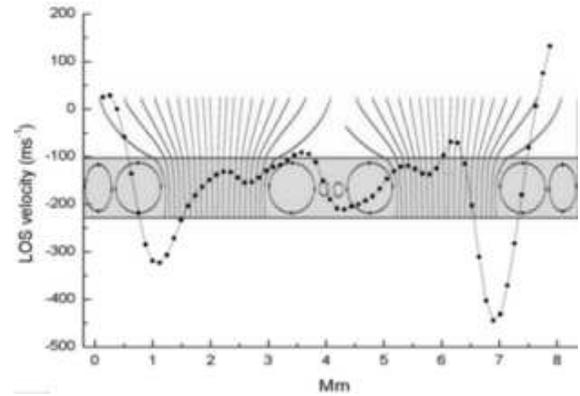}
\caption{Schematic cross-section of the analyzed structure. In the background a sketch of the flows around the pore: the model is obtained by considering  two magnetic flux tubes, each derived from the simulation by Hulburt et al. \cite{hulburt}. Superimposed is the computed mean LoS velocity, averaged along the pore (dot+spline).}
\label{figure10}
\end{figure}

A pore with a light bridge (AR10812) was observed at high spatial and spectral resolution. 
From MDI/SOHO magnetograms and continuum images, we established that the observed region, initially composed of two structures with the same polarity, 
disappears three days after our observation. Such a topology allows to relate the observed light bridge properties to  
the contiguity of two flux tubes. 
With the aim to investigate the photospheric environment of the pore and 
the nature of the bright structure inside it, we computed the intensity and photospheric LoS velocity maps. 
In particular, we analyzed the
intensity maps provided by the broad-band channel of IBIS and by the red wing of 
Ca~II~854.2~nm line, and the corresponding  Fe~I~709.0~nm line LoS velocity fields. \\
Our main conclusions are:
\begin{itemize}
\item The pore is characterized 
by a downward average velocity of $\sim -200~m\cdot s^{-1}$ in the umbra
and of $\simeq -350~m\cdot s^{-1}$ in the surrounding annular region. 
The presence of downflows around magnetic structures has been predicted by
numerical models (e.g.\citet{steiner, hulburt}) and reported in recent observations (e.g. \citet{keil,san}).
\item The time-slice of the 
annular region, irregular in shape and intermittent in time, shows that recurrent strong downflows are present in the boundary of the pore, 
resulting in persistent downflows in the averaged LoS velocity map.
\item An analysis of the intensity and velocity maps and of the relative profiles, calculated along cuts perpendicular 
to the axis of the light bridge,
reveals the presence of elongated structures, showing a kind of reversing in intensity and velocity. More in detail, 
in intensity images we observe a
narrow central dark lane running along the axis of the light bridge. This structure, in our best highest resolution LoS
velocity images, matches to a weak upflow, around $50\div 100~m\cdot s^{-1}$, flanked by two downflows, around $-(200\div 300)~m\cdot s^{-1}$. 
The topology of such a velocity structure resembles a convective roll and may indicate a modification of  
the photospheric convective flows. 
\item We present a semi-analytical model for the light bridge, in which 
we consider a quiet, field free region trapped by two magnetic walls, able to qualitatively reproduce the observed 
intensity behavior inside the light bridge.  
\end {itemize}
By considering the evolution of the analyzed region and taking into account a simulation by \cite{hulburt}, we 
model the observed pore (Fig.\ref{figure10}) as resulting from the merging of two separate magnetic structures with the same polarity both surrounded 
by downward flows. These downflow structures persist in the quenching region, where the convection results strongly modified by 
the presence of the magnetic field, and in the annular region surrounding the pore.\\
This interpretation is supported by the observed time averaged LoS velocity structure reported in Fig.\ref{figure7}. 
This LoS velocity profile is calculated along a cut orthogonal to the light bridge and crosswise to the pore.
Our scheme is analogous to that reported by \cite{jurcak} to explain the magnetic 
canopy above light bridges.\\
As final remark, since bright features inside pores and sunspots show a large {\it zoology}, it should be clear 
that further work is needed on this subject, including a deep investigation of how the magnetic field modifies the local photosphere and
chromosphere.

\begin{acknowledgements}
      The authors thank DST/NSO staff for the efficient support in the observations and in particular we
	  are grateful to M. Bradford, D. Gilliam and J. Helrod. We thank K. Janssen for the IBIS data reduction pipeline and M. Sobotka
	  for useful discussions. This work was partially supported by the MAE Spettro-Polarimetria Solare Bidimensionale 
	  research project and by Regione Lazio CVS (Centro per lo studio della variabilità del Sole) PhD grant.
\end{acknowledgements}

\end{document}